\documentclass{article}
\usepackage{graphicx}  
\usepackage{amsmath}   
\usepackage{amssymb}   
\usepackage{bm} 
\usepackage{dcolumn}
\usepackage{color}
\usepackage{mathrsfs}
\usepackage{amsfonts}
\usepackage{varioref}
\usepackage{slashed}
\usepackage[belowskip=-15pt,aboveskip=0pt]{caption}
\usepackage[margin=1.45in]{geometry}
\RequirePackage[colorlinks,citecolor=blue,urlcolor=magenta,linkcolor=blue]{hyperref}
\def\be{\begin{equation}}
\def\ee{\end{equation}}

\labelformat{section}{Section #1} 
\labelformat{subsection}{Section #1} 
\labelformat{subsubsection}{Section #1}
\labelformat{subsubsubsection}{Section #1}
\labelformat{equation}{Eq.~(#1)} 
\labelformat{figure}{Fig.~#1} 
\labelformat{subfigure}{Fig.~\thefigure#1} 
\labelformat{table}{Tab.~#1} 
\labelformat{appendix}{Appendix #1}

\title{\bf Schwinger effect and a uniformly accelerated observer
}
\author{Shagun Kaushal\footnote{2018phz0006@iitrpr.ac.in}\\\small{Department of Physics, Indian Institute of Technology Ropar, Rupnagar, Punjab 140 001, India}\\}
\date{ }

\begin{document}
\maketitle
\begin{abstract}
\noindent
This article investigates the Schwinger effect for fermions with background electric and magnetic fields of constant strengths from the point of view  of a uniformly accelerated  or the Rindler observer. The Dirac
equation is solved in a closed form, and the field quantisation in the $(3+1)$-dimensional Rindler spacetime is performed. The orthonormal local in and out modes for the causally disconnected right and left wedges and the Bogoliubov relations between them are obtained. Next, the global modes are constructed to cover
the whole spacetime, and the Bogoliubov relationship between the local and global operators is found. Using them the squeezed state expansion of the global vacuum in terms of local states is acquired and accordingly, the spectra of created particles is found. Clearly, there are two
sources of particle creation in this scenario -- the Schwinger as well as the Unruh effects. Our chief aim is to investigate the role of the strength of the background electromagnetic fields on the spectra of created particles. We also  discuss very briefly some possible implication of this
result in the context of quantum entanglement.
\end{abstract}
\newpage
\tableofcontents
\section{Introduction}
\label{introduction}
Particle creation in curved or non-trivial backgrounds is an exciting phenomenon \cite{Parker:2009uva}. In such scenarios, the vacuum may spontaneously break down due to
quantum fluctuations \cite{ quanfluc, Karabali:2019ucc, UW}, and virtual pairs can be separated either by the energy of the
fields or the causality of spacetime. These quantum field theoretical effects cause black holes to create and emit particles, known as the Hawking radiation \cite{Hawking, Hawking1}. In the cosmological scenario on the other hand, pair creation occurs due to the expansion of the spacetime \cite{dS, dS1}.
A uniformly accelerated detector moving in flat spacetime perceives the Minkowski vacuum to be thermally populated at temperature $T = a/2\pi$ where $a$ is the acceleration parameter \cite{dS1, Unruh, Dirac1, rindler1, rindler2, leftright, unruhdewitter1, unruhdewitter2, unruh_exp},  known as the Unruh effect.  
The vacuum of a charged quantum field  is also unstable in the presence of a background electric field, which leads to pair production, known as the Schwinger effect \cite{Schwinger}.

In quantum electrodynamics, a magnetic field alone cannot give rise to pair creation but can affect its rate if a background electric field is also present. The effect of a magnetic field on the Schwinger effect for a complex scalar field was discussed in \cite{HSSS} and was shown that a sufficiently high strength of the magnetic field stabilises the vacuum. An interesting question here is how a uniformly accelerated observer would see these oppositely moving particles created by the electric field? And what will be the role of the background magnetic field in this context?
A similar analysis was done for fermions  in cosmological de Sitter spacetime in the presence of  primordial electromagnetic fields~\cite{SSSS}. There are two sources of particle creation in this scenario : the background electromagnetic field and the expansion of the spacetime and it was shown that the magnetic field does not  at all affect the pair creation due to the expansion of spacetime. Thus it seems interesting to ask : how the electromagnetic field will affect the pair creation due to the Unruh effect, for a charged quantum field? Precisely, there will be two sources of particle creation here. One will be the Schwinger effect in a given Rindler wedge. The other source certainly correspond to the existence of a global vacuum state. We wish to incorporate {\it both} of these in our composite scenario. 

In \cite{SHG, SHU}, the Schwinger effect is studied in the (anti-)de Sitter spaces and black hole backgounds, representing a unified picture of the Schwinger effect and the Hawking radiation. The Schwinger effect from the near-extremal black hole is observed to be a product of the AdS$_2$ Schwinger effect and a correction due to the Hawking radiation from non-extremality. 
The quantization of a charged scalar field in the Rindler spacetime with a constant strength background electromagnetic field might have relevance in the non-extremal astrophysical black hole spacetimes. Also in particular, it may give some physical insight into the dynamics of charge quantum fields in the near horizon geometries of charged black hole spacetimes. 

In \cite{Rindler1+1} the quantisation of a complex scalar field interacting with a constant background electric field in the $(1+1)$-Rindler spacetime was performed
and an expression of the vacuum decay rate was found.
 The main characteristic of this problem is that it involves two accelerations: the acceleration of the Rindler observer and the acceleration due to the non-zero electric field. Firstly, the Schwinger vacuum decay rate expression is established in the Minkowski spacetime. Next, the construction of Unruh modes is discussed to connect the Minkowski and Rindler modes. Using the Unruh modes, the mean number density of particles and antiparticles observed by an accelerated observer with respect to the Minkowski vacuum is computed. It has been observed that the particles and antiparticles are not equally distributed in the particular Rindler wedges, which leads to charge polarization. However, the total charge is always conserved, i.e., the sum of the total charges in the R and L Rindler wedges. They also talked about the difference in quantization in the Minkowski space and the Rindler wedges. 

A similar, perhaps more realistic, scenario is considered here for the charged fermionic field with the non-zero background electromagnetic fields in $(3+1)$-dimensional Rindler spacetime. Along with two sources of acceleration, we have a magnetic field also. In addition, we shall also carry out the construction of the global modes existing in both the causally disconnected wedges. Even though we consider global modes, the Bogoliubov transformations we obtained are for local wedges. We constructed the global Minkowski vacuum using the Bogoliubov transformations between the local modes.
An exciting outcome of the global modes  is the emergence of entanglement between the left and right wedges, e.g.~\cite{FuentesSchuller:2004xp, Yi Ling, RL}. 
Entanglement in the context of black hole or the Rindler spacetime receiving attention for the past few years \cite{EE_1, EE_2}. Various measures to quantify entanglement are discussed in 
\cite{NielsenChuang, Agarwal:2016cir, jmath, Plenio:2007zz, Monogamy, Zyczkowski:1998yd, Martin, Vidal:2002zz, wang, Vidal:1998re, Horodecki:2009zz}.
Entanglement is an observer-dependent quantity in non-inertial frames, shown explicitly by \cite{FuentesSchuller:2004xp, Yi Ling, MShamirzai, MartinMartinez:2010ar, unruh4, unruhgaussian}, it originates due to the appearance of the event horizon that results in a loss of information for the non-inertial observer. For scalars and Dirac fields, the degradation of entanglement occurs from the perspective of a uniformly accelerated observer \cite{MShamirzai, Tel,Dirac2}. Whereas aspects of entanglement for the particles created by Schwinger effect are also studied in \cite{Schwinger1, Schwinger2, Dai:2019nzv, Li:2018twv, UnSc1}. Thus it would also be interesting to see whether the entanglement between the particles created from the Schwinger effect would also depend on the motion of the observer, in the mixed kind of framework  we wish to investigate. 

 In \ref{section : S2}, the Dirac field is quantised in the right (R) and left (L) Rindler wedges explicitly, with respect to the orthonormal modes found in closed form. The Bogoliubov relations between the local modes due to the background electriomagnetic field are shown explicitly. In \ref{S3}, the global modes existing on both the wedges  are constructed. The relation between the local and global creation and annihilation operators is obtained, which incorporates the two Bogoliubov transformations, one due to the Schwinger and the other due to the Unruh effect. 
In \ref{S4} the global vacuum number density and logarithmic negativity to look at the vacuum instability and entanglement properties between the created particles are computed.Finally, we concluded in \ref{S5}. Computational detail  are provided in the appendices. A possible further extension of this work can be the study entanglement properties for a composite state constructed by two fields, discussed earlier by \cite{HSSS, SSSS, degradation:2015, degradation:2015n,  Global_3, SHN:2020, Fuentes:2010dt, BKAY:2017, nper:2011, KAKAY-2021,  Rindler_entanglement1}. 

We shall work with the mostly positive signature of the metric in $(3+1)$-dimensions and will set $c=k_B=\hbar=1$ throughout.
\section{ The Dirac modes}\label{section : S2}
\begin{figure}[!ht]
\begin{center}
     \includegraphics[scale=.42]{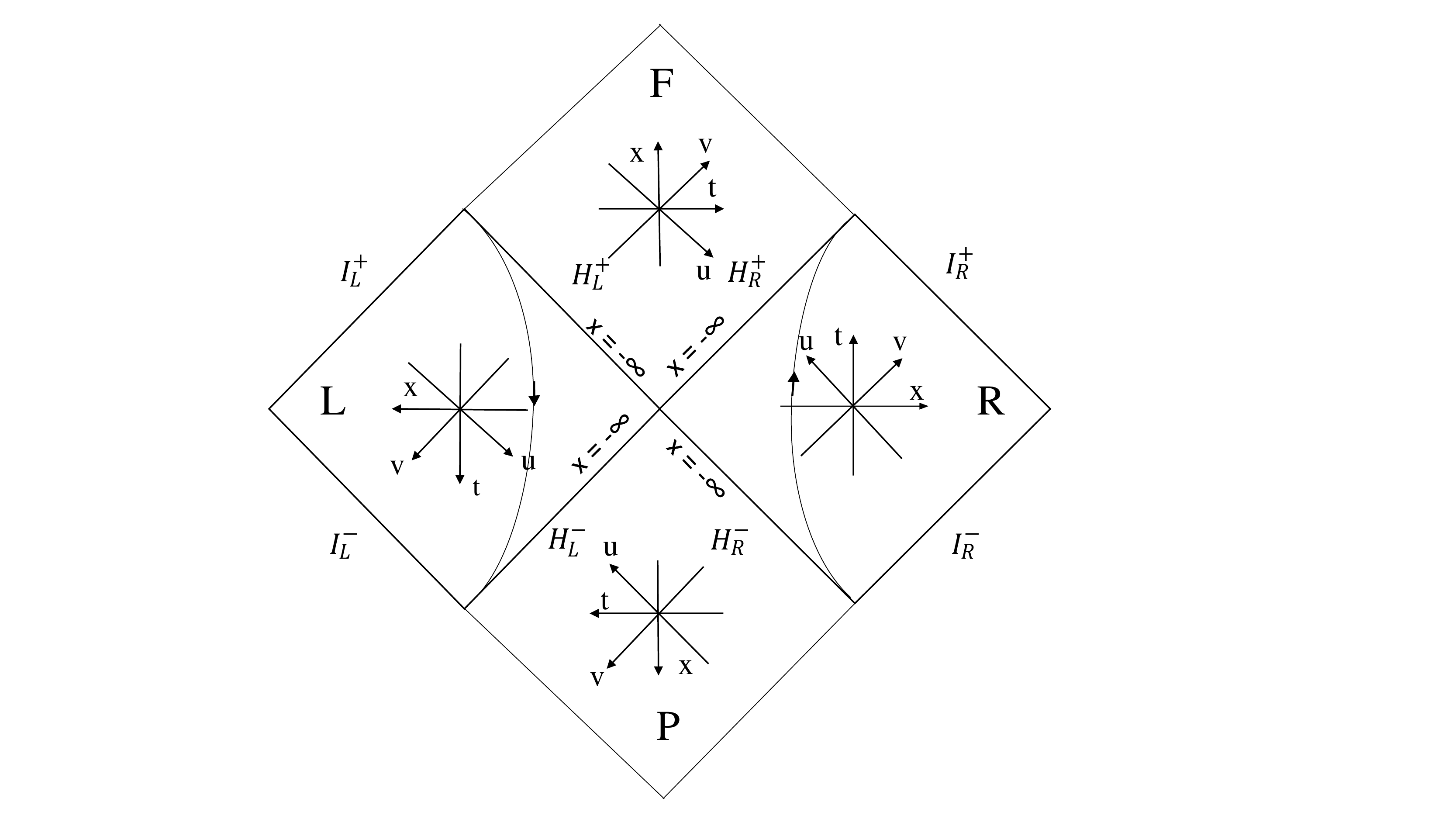}
    \caption{The four Rindler patches $R$, $L$, $P$, and $F$, with their coordinates. Here $\mathit{I^-_{R,L}}$ and $\mathit{I^+_{R,L}}$ are the past and future null infinities, whereas $\mathit{H^-_{R,L}}$ and $\mathit{H^+_{R,L}}$ are the past and future horizons. The hyperbolic curves represent the trajectories of particles, whereas $u$ and $v$ represent the lightlike coordinates.}
    \label{diagram}
    \end{center}
\end{figure}

The Rindler coordinate transformations divide the Minkowski space into two patches, denoted hereafter
by the labels $R$ and $L$. On each of these patches the coordinate
transformations between Minkowski $\left(\tau, \rho, y, z\right)$, and Rindler $\left(t, x, y,z\right)$ coordinates are given as
\begin{eqnarray}
\label{transformation}
    \tau=\frac{1}{a}e^{ax_R}\sinh a t_R,\; \rho=\frac{1}{a}e^{ax_R}\cosh a t_R\; {\rm\left(R\right)};\;
     \tau=\frac{1}{a}e^{ax_L}\sinh a t_L,\; \rho=-\frac{1}{a}e^{ax_L}\cosh a t_L\; \rm{\left(L\right)}
\end{eqnarray}
here for each of these quadrants, the respective Rindler coordinates run from $-\infty$ to $\infty$ shown in \ref{diagram}. On R and L, the vector field $\partial_t$ is timelike.
The world lines of uniformly accelerated observers in the Minkowski coordinates correspond to hyperbolas to the left and right of the origin, which are
bounded by lightlike asymptotes constituting the Rindler horizon, so the two Rindler regions are causally disconnected from each other. An observer undergoing uniform acceleration remains constrained to either the Rindler region $R$ or $L$ and has no access to the other sector. In \ref{diagram}, $\mathit{I^-_{R,L}}$ and $\mathit{I^+_{R,L}}$ are the past and future null infinities, whereas $\mathit{H^-_{R,L}}$ and $\mathit{H^+_{R,L}}$ are the past and future horizons whereas $u$ and $v$ are lightlike coordinates defined as $u=t-x$ and $v=t+x$. Under the transformation \ref{transformation}, the line element takes the form
\begin{equation}
\label{metric1}
ds^2=e^{2ax}(-dt^2+dx^2)+dy^2+dz^2,
\end{equation}
where $a$ is the acceleration parameter, the metric is the (3+1)-Rindler spacetime metric.
 Let us now focus on the fermionic field theory coupled to external or background electromagnetic fields in the four-dimensional Rindler spacetime. \\
The Dirac equation for the $\psi$ in curved spacetime is given as \cite{peskin}
\begin{equation}
\label{dirac1}
 (i\gamma^{\mu}D_{\mu}-m)\psi(x) = 0
\end{equation}
here the gauge cum spin covariant derivative is defined as
\begin{eqnarray}
\label{spin_conn}
D_\mu=\partial_\mu+i q A_\mu+\Gamma_\mu
\end{eqnarray}
ensuring the local gauge symmetry and the general covariance.
Here $\Gamma_\mu$ is the spin connection and its only non-zero component is $\Gamma_0$ written as
\begin{equation}
\Gamma_0=\frac{a}{2}\gamma^{(1)}\gamma^{(0)},    
\end{equation}
here $\gamma^{(a)'}s$ are the flat spacetime gamma matrices.
Next, we introduce the tetrads $e^\mu_a$ ($a,b,c...= 0,1,2,3$ are indices for the local Lorentz transformation and the Greek indices $\mu,\nu$···are for spacetime), the tetrad field is related to the metric in curved spacetime with the help of the four-dimensional Minkowski metric as

\begin{eqnarray}
\label{tetrad}
g_{\mu\nu}=e^{a}_\mu e^{b}_\nu \eta_{ab}. 
\end{eqnarray}
Following from \ref{tetrad} and \ref{metric1} we choose the tetrads for the Rindler metric \ref{metric1} as
\begin{equation}
\label{tetrad1}
e^{\mu}_{a}=\text{diag}(e^{-ax},e^{-ax},1,1)
\end{equation}
Defining a new variable $\widetilde{\psi}=e^{\frac{a x}{2}} \psi$ in equation \ref{dirac1}, it becomes
\begin{eqnarray}
\label{dirac2}
(i e^\mu_a \gamma^{(a)}\partial_\mu-q e^\mu_a \gamma^{(a)}A_\mu-m)\Tilde{\psi}(x)=0
\end{eqnarray}
Substituting next
\begin{equation}
\label{de0}
\widetilde{\psi}(x)= (i e^\mu_a \gamma^{(a)}\partial_\mu-q A_\mu e^\mu_a \gamma^{(a)}+m)\chi(x)
\end{equation}
 in \ref{dirac2} gives
\begin{equation}
    \label{dirac3}
 \bigg(\frac{1}{e^{2ax}}\Big((\partial_t+i q A_0)^2+a \partial_{x}-\partial_{x}^{2}\Big) -\partial^2_y-(\partial_z+i q A_3)^2+\frac{\gamma^{(1)} \gamma^{(0)}}{e^{2 a x}}\big(a \partial_t-i q \partial_x A_0\big)-i q \partial_{y}A_3 \gamma^{(2)}\gamma^{(3)}-m^2\bigg)\chi(x)=0
\end{equation}
We choose the gauge which gives us constant electric and  magnetic fields along $x-$axis as
\begin{equation}
    \label{gauge}
    A_\mu \equiv \frac{E  e^{2ax}}{2 a} \delta^t_{\mu} + B y \delta^z_{\mu}
\end{equation}
where $E, \;B $ and $a$ are the constants. We consider the ansatz $\chi(x)=e^{-i\omega t} e^{ik_z z}\zeta_s(x,y) \epsilon_s$ (no sum on $s$) in \ref{dirac3}, we have
\begin{equation}
\label{dirac4}
    \bigg(-\frac{1}{e^{2ax}}\big( \omega - \frac{q E e^{2ax}}{2 a}\big)^2+\frac{a}{e^{2 a x}} \partial_{x}-\frac{1}{e^{2 a x}}\partial_{x}^{2}-\partial^2_y+(k_z+ q B y)^2-i\frac{ \gamma^{(1)} \gamma^{(0)}}{e^{2ax}}(a \omega+\frac{ q E e^{2ax} }{ 2})-i q B \gamma^{(2)}\gamma^{(3)}-m^2\bigg)\zeta_s(x,y) \epsilon_s=0 
\end{equation}
where $\epsilon_s$ are the simultaneous eigenvectors of $\gamma^{(1)}\gamma^{(0)}$ and $\gamma^{(2)}\gamma^{(3)}$, such that it has the following eigenvalue equations with $\gamma^{(1)}\gamma^{(0)}$ and $\gamma^{(2)}\gamma^{(3)}$, $\gamma^{(1)}\gamma^{(0)}\epsilon_1=-\epsilon_1$, $\gamma^{(1)}\gamma^{(0)}\epsilon_2=-\epsilon_2$, $\gamma^{(1)}\gamma^{(0)}\epsilon_3=\epsilon_3 $, $\gamma^{(1)}\gamma^{(0)}\epsilon_4=\epsilon_4$, $\gamma^{(2)}\gamma^{(3)}\epsilon_1=-i \epsilon_1$, $\gamma^{(2)}\gamma^{(3)}\epsilon_2=i \epsilon_2$, $\gamma^{(2)}\gamma^{(3)}\epsilon_3=-i \epsilon_3$ and $\gamma^{(2)}\gamma^{(3)}\epsilon_4=i \epsilon_4$ whereas the explicit form of $\epsilon_s$ are as follows
\begin{center}$\epsilon_1= \frac{1}{\sqrt{2}}\left( {\begin{array}{cccc}
  0\\
 0\\
1\\
   1\\
  \end{array} } \right)$,  
  $\epsilon_2= \frac{1}{\sqrt{2}}\left( {\begin{array}{cccc}
  -1\\
 1\\
0\\
   0\\
  \end{array} } \right)$,
$\epsilon_3=\frac{1}{\sqrt{2}} \left( {\begin{array}{cccc}
  0\\
 0\\
-1\\
   1\\
  \end{array} } \right) $and $\epsilon_4= \frac{1}{\sqrt{2}}\left( {\begin{array}{cccc}
  1\\
 1\\
0\\
   0\\
  \end{array} } \right)$.
 \end{center}
By using the eigenvalue equations and separation of variables as done in \cite{SSSS}, it gives us two differential equations as follows
 \begin{equation}
 \label{rho1}
     \bigg(\partial^2_{x}-a \partial_x+\omega^2-\frac{q E \omega}{a}e^{2 a x }-i \omega a-\bigg(\frac{q E e^{2 a x}}{2 a}\bigg)^2 -\frac{i q E}{2}e^{2 a x}+(m^2+S_s) e^{2 a x}\bigg)\zeta_s(x)=0
 \end{equation}
and 
\begin{equation}
    \label{y1}
    \bigg(\partial^2_y-(k_z+qBy)^2+q B-S_s\bigg)H_s(y)=0
\end{equation}
where
\begin{equation}
\label{seapration}
  S_1=-2 n_L q B \;\text{and}\; S_2=-(2 n_L+1) q B  
\end{equation}are the separation constant and $n_L$ corresponds to the Landau level. 
The general solution of \ref{rho1} for $s=1$, i.e. $\zeta_1(x)$ are $e^{a x}e^{-\frac{i q E e^{2 a x}}{4 a^2}} e^{i \omega x} U(\lambda_1, \nu, \xi )$ and  $e^{a x} e^{-\frac{i q E e^{2 a x} }{4 a^2}}  e^{i \omega x} L(-\lambda_1, \nu-1, \xi) $, where $U$ and $L$ are the confluent hypergeometric and the generalized Laguerre functions respectively whose explicit form are given in \cite{AS},
whereas solution of \ref{y1}
\begin{equation}
    \label{soly1}
    H_s(y)=\left(\frac{\sqrt{qB}}{2^{n+1}\sqrt{\pi}(n+1)!}\right)^{1/2}e^{-\Tilde{y}^2/2}\mathcal{H}_{n}(\Tilde{y})=h_{n}(\Tilde{y})~({\rm say})
\end{equation}
here $\lambda_1$, $\lambda_2$ and $\nu$ are the parameters defined as

\begin{equation}
\label{coeff}
\lambda_1=\lambda_3=\frac{i(m^2 + S_1)+2 q E}{2 q E}, \;
\lambda_2=\lambda_4=\frac{i(m^2 + S_2)+2 q E}{2 q E}, \;
\nu=\frac{3}{2}+\frac{i \omega}{a}
\end{equation}
 whereas the variables $\Tilde{y}$ and $\xi$ are defined as
\begin{equation}
\label{ell}
\xi=-\frac{i q E e^{2a x}}{2 a^2},\;
    \Tilde{y}=\left(\sqrt{qB}y+\frac{k_z}{\sqrt{qB}}\right) 
\end{equation}
Let us now find out the $\mathit{in}$ modes for $R$ wedge, where at $x \to \infty$ and  $x \to -\infty$, that corresponds to $\mathit{I^-_R}$ and $\mathit{H^-_R}$ respectively \ref{diagram}. Mode emerging from $\mathit{H}^-_R$ is moving towards $\mathit{I^+_R}$ and the relevant part of mode proportional to $e^{- i \omega u}$

\begin{center}
   $ \zeta_s(x) \sim e^{a x} e^{-\frac{i q E e^{2 a x}}{4 a^2}} e^{-i \omega (t-x)} \xi^{-\lambda_s},\;s=1,2 $ 
\end{center}

Similarly, modes emerging from $\mathit{I^-_R}$ are moving towards $\mathit{H^+_R}$ and the relevant part of mode is proportional to $e^{- i \omega v}$

\begin{center}
      $ \zeta_s(x) \sim e^{ax} e^{\frac{i q E e^{2 a x} }{4 a^2}} e^{-i \omega (t+x)} \frac{\Gamma(-\lambda_s^*+ \nu^*)}{\Gamma(\nu^*) \Gamma(-\lambda_s^* + 1)},\;s=1,2  $  
\end{center}
Putting these together we have four modes from which two corresponds to $\mathit{I^-_R}$ and two to $\mathit{H^-_R}$ written as
\begin{eqnarray}
\label{chi}
\chi(x)_{\mathit{H^-_R}_{,s}}&=&e^{-i\omega( t-x)} e^{i k_z z} e^{a x} e^{-\frac{i q E e^{2 a x}}{4 a^2}} U(\lambda_s, \nu, \xi) H_s(y) \epsilon_s\\
\label{chi1}
 \chi(x)_{\mathit{I^-_R}_{,s}}&=&e^{-i\omega ( t+x)} e^{i k_z z} e^{a x} e^{\frac{i q E e^{2 a x}}{4 a^2}} (L(-\lambda_s, \nu-1, \xi))^*  H_s(y) \epsilon_s
\end{eqnarray}
here $s=1,2$ in \ref{chi} and \ref{chi1}.
\subsection{Quantization on the right wedge (R)}
For computing full modes we need to substitute $\chi(x)$ in \ref{de0} and then using the definition $\psi=e^{-\frac{a x}{2}}\Tilde{\psi}$ the final particle in modes are given as
\begin{equation}
    \label{mode1}
    U_{s,n}(x)_{\mathit{H^-_R}}=\frac{ e^{-\frac{a x}{2}} }{N_s} (ie^\mu_a \gamma^{(a)}\partial_\mu-q A_\mu e^\mu_a \gamma^{(a)}+m) e^{-i\omega( t-x)} e^{i k_z z} e^{a x} e^{-\frac{i q E e^{2 a x}}{4 a^2}} U(\lambda_s, \nu, \xi) H_s(y) \epsilon_s,
\end{equation}
\begin{equation}
    \label{mode2}
    U_{s,n}(x)_{\mathit{I^-_R}}=\frac{e^{-\frac{a x}{2}}}{M_s} (i e^\mu_a \gamma^{(a)}\partial_\mu-q A_\mu e^\mu_a \gamma^{(a)}+m) e^{-i\omega ( t+x)} e^{i k_z z} e^{a x} e^{\frac{i q E e^{2 a x}}{4 a^2}} (L(-\lambda_s, \nu-1, \xi))^*  H_s(y) \epsilon_s,
\end{equation}
here $s=1, 2$ for \ref{mode1} and \ref{mode2} and the parameter $\xi$ is defined in \ref{ell}, they are the positive-frequency modes with respect to a future-directed time like Killing vector $\partial_t$, whereas negative energy modes are given as
\begin{equation}
    \label{mode1a}
    V_{s,n}(x)_{\mathit{H^-_R}}=\frac{ e^{-\frac{a x}{2}} }{P_s} (ie^\mu_a \gamma^{(a)}\partial_\mu-q A_\mu e^\mu_a \gamma^{(a)}+m) e^{i\omega(t-x)}e^{ax}e^{i k_z z}e^{-\frac{i q Ee^{2ax}}{4a^2}}e^{\xi}U(\nu-\lambda_s,\nu,\xi)H_s(y)\epsilon_s,
\end{equation}
\begin{equation}
    \label{mode2a}
    V_{s,n}(x)_{\mathit{I^-_R}}=\frac{e^{-\frac{a x}{2}}}{R_s} (i e^\mu_a \gamma^{(a)}\partial_\mu-q A_\mu e^\mu_a \gamma^{(a)}+m) e^{i\omega(t+x)}e^{ax}e^{i k_z z}e^{\frac{i q Ee^{2ax}}{4a^2}}\xi^{1-\nu^*}(L(\nu-\lambda_s-1,1-\nu,\xi))^*H_s(y)\epsilon_s,
\end{equation}
here $s=3, 4$ for \ref{mode1a} and \ref{mode2a}, and $N_s$, $M_s$, $P_s$ and $R_s$ are the normalization constants obtained by normalizing modes at constant $u$ and $v$ surfaces, which are shown in the \ref{A} explicitly. Orthonormality relations of these modes are given by
\begin{eqnarray}
&&(U_{s,n}(x)_{\mathit{H^-_R}, \mathit{I^-_R}},U_{s^{\prime},n^{\prime}}(x)_{\mathit{H^-_R, \mathit{I^-_R}}})= (V_{s,n}(x)_{\mathit{H^-_R}, \mathit{I^-_R}},V_{s^,n^{\prime}}(x)_{\mathit{H^-_R}, \mathit{I^-_R}} )= \delta(k_z-k'_z)\delta(\omega-\omega ')\delta_{n n^{\prime}}\delta_{ss^{\prime}}
\end{eqnarray}
Next, we have computed another set of orthonormal outmodes corresponding to the regions $\mathit{I^+_R}$ and $\mathit{H^+_R}$ of \ref{diagram}. Since the $\mathit{in}$-basis functions contain a particle plus antiparticle in the future, they are not useful to describe quantization in terms of single quanta in the future. Their time reversed versions are given by following definition $U_{s,n}(t,\vec{x})_{\mathit{I^+_R},\mathit{H^+_R}}=(U_{s,n}(-t,\vec{x})_{\mathit{I^-_R},\mathit{H^-_R}})^{*}$ analogous to \cite{out_def, out_def1}, in which they have used the same definition to compute out modes for scalar field in the $(1+1)$-Rindler spacetime, similarly for negative frequency $\mathit{out}$ modes we have $V_{s,n}(t,\vec{x})_{\mathit{I^+_R},\mathit{H^+_R}}=(V_{s,n}(-t,\vec{x})_{\mathit{I^-_R},\mathit{H^-_R}})^{*}$ (explicit form of the modes and the calculation of normalization constants are shown in  the \ref{A}). We now make the field quantization on $R$ in terms of the modes on them as follows
\begin{align}
\label{FQR1}
\psi_R(x) &= \sum_{n; s}\int\frac{d\omega d k_z}{2\pi } \Bigg[a_{\rm }(\omega,k_z,s,n)_{\mathit{H^-_R}}U_{s,n}(x;\omega,k_z)_{\mathit{H^-_R}}+b^{\dagger}_{\rm}(\omega,k_z,s,n)_{\mathit{H^-_R}}V_{s,n}^*(x;\omega,k_z)_{\mathit{H^-_R}}\\ &+ a_{\rm}(\omega,k_z,s,n)_{\mathit{I^-_R}}U_{s,n}(x;\omega,k_z)_{\mathit{I^-_R}}+b^{\dagger}_{\rm }(\omega,k_z,s,n)_{\mathit{I^-_R}}V_{s,n}^*(x;\omega,k_z)_{\mathit{I^-_R}}\Bigg] \\
&=\sum_{n; s}\int\frac{d\omega d k_z}{2\pi } \Bigg[a_{\rm}(\omega,k_z,s,n)_{\mathit{H^+_R}}U_{s,n}(x;\omega,k_z)_{\mathit{H^+_R}}+b^{\dagger}_{\rm }(\omega,k_z,s,n)_{\mathit{H^+_R}}V_{s,n}^*(x;\omega,k_z)_{\mathit{H^+_R}}\\ &+a_{\rm }(\omega,k_z,s,n)_{\mathit{I^+_R}}U_{s,n}(x;\omega,k_z)_{\mathit{I^+_R}}+b^{\dagger}_{\rm}(\omega,k_z,s,n)_{\mathit{I^+_R}}V_{s,n}^*(x;\omega,k_z)_{\mathit{I^+_R}}\Bigg]
\end{align}
here the creation and annihilation operators are assumed to satisfy the usual canonical anti-commutation relations.
Using the relation between confluent hypergeometric functions \cite{AS}, we can write the Bogoliubov relation between $\mathit{in}$ and $\mathit{out}$ modes as
\begin{eqnarray}
    \label{BTM}
     U_{s,n}(x)_{\mathit{H^-_R}}=\alpha_{s}^{*}  U_{s,n}(x)_{\mathit{I^+_R}}+\beta_s^{*}  ( V_{s,n}(x)_{\mathit{I^+_R}})^*, \nonumber\\
     V_{s,n}(x)_{\mathit{H^-_R}}=\alpha_{s}^{*}  V_{s,n}(x)_{\mathit{I^+_R}}+\beta_s^{*}  ( U_{s,n}(x)_{\mathit{I^+_R}})^*
\end{eqnarray}
here $s=1,2$ in \ref{BTM} and $\alpha_s$ and $\beta_s$ are the Bogoliubov coefficients given as
\begin{align}
\label{Bcoeff}
\alpha_s = \frac{N_s \Gamma(1-\lambda_s)\sin\pi(\lambda_s-\nu)}{M_s\sin \pi  \nu },\quad
\beta_s = \frac{N_s \sin \pi  \lambda_s \Gamma (\nu -\lambda_s )}{R_s\sin \pi  \nu  }
\end{align}
whereas the Bogoliubov transformation between the creation and annihilation operators are given as
\begin{eqnarray}
\label{Bogoin_out}
&&a_{\rm }(\omega,k_z,s,n)_{\mathit{H^-_R}}\;=\;\alpha_s a_{\rm }(\omega,k_z,s,n)_{\mathit{I^+_R}}-\beta_{s}^{*} b_{\rm }^{\dagger}(-\omega,-k_z,s,n)_{\mathit{I^+_R}}, \\
&&b_{\rm}(\omega,k_z,s,n)_{\mathit{H^-_R}}\;=\;\alpha_s b_{\rm }(\omega,k_z,s,n)_{\mathit{I^+_R}}+\beta_{s}^{*} a^{\dagger}_{\rm }(-\omega,-k_z,s,n)_{\mathit{I^+_R}}, 
\end{eqnarray}
The canonical anti-commutation relations ensure,
$\lvert\alpha_s\rvert^2+\lvert\beta_s\rvert^2=1 $. The coefficient $\beta_s$ is responsible for pair production, and the quantity $\lvert \beta_s \rvert^2$ is the mean number density of particles
\begin{equation}
    \label{beta2}
\lvert\beta_s\rvert^2= \frac{\sinh^3{\pi \Delta}}{e^{\pi \Delta } \cosh^3{\pi(\Delta-\frac{\omega}{a})}+\sinh^3{\pi \Delta}} 
\end{equation}
here the parameter $\Delta$ is defined as,
\begin{equation}
\label{delta}
   \Delta = \text{Im}{(\lambda)}=\frac{m^2+S_s}{2qE} 
\end{equation}
and $\omega>0$. In \ref{beta2}, $\lvert\beta_s\rvert^2$ is independent of momentum. For $\Delta \to \infty$, that corresponds to zero electric field $(E \to 0)$ or zero electric charge $(q \to 0)$ or large magnetic field $(B\to \infty)$ the number density for local vacuum leads to zero, $\lvert\beta_s\rvert^2 \to 0$  and this behaviour is similar to the usual Minkowski vacuum in the presence of background electromagnetic field \cite{HSSS} and the result obtained for scalar field with background electric field in the Rindler spacetime \cite{UnSc1}. Whereas for the non-zero strength of the electromagnetic field, the number density also depends on the acceleration of the non-inertial observer. 
\subsection{Quantization on the left wedge (L)}
The field equations and their solutions are the same on the left and right wedges. The only difference between $R$ and $L$ wedge is that on $L$, the vector field $\partial_t$ points near the past whereas $\partial_{t_L}=-\partial_t$ plays the role of future-directed Killing vector as explained in \cite{leftright}, which implies that the sign of the charges and the $\mathit{in}$ and $\mathit{out}$ labels have to interchange concerning their values on the $R$ wedge. Hence, the complete set of $\mathit{in}$ modes for $L$ wedge are given as
\begin{equation}
\label{fullmode1L}
    U_{s,n}(x_L)_{\mathit{H^-_L}}=\frac{e^{-\frac{a x_L}{2}}}{N_s} (ie^\mu_a \gamma^{(a)}\partial_\mu-q A_\mu e^\mu_a \gamma^{(a)}+m)e^{-i \omega (t_L+x_L)} e^{i k_z z} e^{a x_L} e^{-\frac{i q E e^{2 a x_L}}{4 a^2}} (U(\lambda_s, \nu, \xi_L))^*H_s(y) \epsilon_s,
\end{equation}
\begin{equation}
\label{fullmode2L}
    U_{s,n}(x_L)_{\mathit{I^-_L}}=\frac{e^{-\frac{a x_L}{2}}}{M_s} (ie^\mu_a \gamma^{(a)}\partial_\mu-q A_\mu e^\mu_a \gamma^{(a)}+m) e^{-i \omega (t_L-x_L)} e^{i k_z z} e^{a x_L} e^{-\frac{i q E e^{2 a x_L}}{4 a^2}} L(-\lambda_s, \nu-1, \xi_L) H_s(y) \epsilon_s,
\end{equation}
\begin{equation}
\label{fullmode1aL}
    V_{s,n}(x_L)_{\mathit{H^-_L}}=\frac{e^{-\frac{a x_L}{2}}}{P_s} (ie^\mu_a \gamma^{(a)}\partial_\mu-q A_\mu e^\mu_a \gamma^{(a)}+m)e^{i \omega (t_L+x_L)} e^{-i k_z z} e^{a x_L} e^{-\frac{i q E e^{2 a x_L}}{4 a^2}} (e^{\xi_L}U(\nu-\lambda_s,\nu,\xi_L))^* H_s(y) \epsilon_s,
\end{equation}
\begin{equation}
\label{fullmode2aL}
    V_{s,n}(x_L)_{\mathit{I^-_L}}=\frac{e^{-\frac{a x_L}{2}}}{R_s} (ie^\mu_a \gamma^{(a)}\partial_\mu-q A_\mu e^\mu_a \gamma^{(a)}+m) e^{i \omega (t_L-x_L)} e^{-i k_z z} e^{a x_L} e^{-\frac{i q E e^{2 a x_L}}{4 a^2}} \xi_L^{1-\nu} L(\nu-\lambda_s-1,1-\nu,\xi_L)  H_s(y) \epsilon_s,
    \end{equation}
where $s=1, 2$ for \ref{fullmode1L}, \ref{fullmode2L} and $s=3, 4$ for \ref{fullmode1aL},  \ref{fullmode2aL}, the parameter $\xi$ is defined by \ref{ell}, and they  are the positive and negative energy modes with respect to $\partial_{t_L}$ respectively. Now, by using the definition for $\mathit{out}$ modes defined in previous section, we can find the $\mathit{out}$ modes for $L$ wedge also, showm in \ref{A} explicitly. Whereas the Bogoliubov transformation as well as coefficients will remain similar as that of $R$.

\section{The global modes and Bogoliubov coefficients}\label{S3}
Since $R$ and $L$ are just two patches of the Minkowski spacetime, they do not cover the whole Minkowski spacetime; therefore, we form global modes to cover the whole Minkowski spacetime. The set of modes on $\mathit{I^-_R} \cup \mathit{H^-_R} $ and $\mathit{I^+_R} \cup \mathit{H^+_R} $ are disconnected in \ref{diagram}, therefore to cover the whole Minkowski spacetime we construct global modes having support in $R \cup L$ using Unruh's prescription as used in \cite{Global_3, Global_1}. For the construction of global $\mathit{in}$ modes, we take the linear combination of the modes on $R$ and $L$ wedges by comparing their asymptotic limit behavior of modes shown explicitly in \ref{B}. According to that the set of global $\mathit{in}$ modes are constructed by the superposition of $U_{1,n}(x)_{\mathit{H^-_R}}, V_{1,n}(x)_{\mathit{H^-_R}}, U_{1,n}(x)_{\mathit{I^-_L}}$ and $V_{1,n}(x)_{\mathit{I^-_L}}$, are as follows
\begin{eqnarray}
\label{Gmode1}
    \phi^{G}_{1}(x)=\frac{1}{\sqrt{2\; \cosh\frac{\omega \pi}{a}}}\big(e^{\frac{\pi \omega}{2a}} U_{1,n}(x)_{\mathit{H^-_R}}+e^{-\frac{\pi \omega}{2a}}V_{1,n}(x)_{\mathit{I^-_L}}\big)\\
    \label{Gmode11}
    \phi^{G}_{2}(x)=\frac{1}{\sqrt{2\; \cosh\frac{\omega \pi}{a}}}\big(e^{\frac{\pi\omega}{2a}} U_{1,n}(x)_{\mathit{I^-_L}}+e^{-\frac{\pi\omega}{2a}}V_{1,n}(x)_{\mathit{H^-_R}}\big)\\
    \label{Gmode111}
     \phi^{G}_{3}(x)=\frac{1}{\sqrt{2\; \cosh\frac{\omega \pi}{a}}}\big(e^{\frac{\pi\omega}{2a}} V_{1,n}(x)_{\mathit{H^-_R}}-e^{-\frac{\pi \omega}{2a}}U_{1,n}(x)_{\mathit{I^-_L}}\big)\\
     \label{Gmode1111}
    \phi^{G}_{4}(x)=\frac{1}{\sqrt{2\; \cosh\frac{\omega \pi}{a}}}\big(e^{\frac{\pi \omega}{2a}}V_{1,n}(x)_{\mathit{I^-_L}}-e^{-\frac{\pi \omega}{2a}}U_{1,n}(x)_{\mathit{H^-_R}}\big)
\end{eqnarray}
Therefore, \ref{Gmode1}, \ref{Gmode11}, \ref{Gmode111} and \ref{Gmode1111} are the global modes in terms of local $\mathit{in}$ modes, whereas the global modes can be obtained in terms of local  $\mathit{out}$ modes by using the local modes Bogoliubov transformation from \ref{BTM} in \ref{Gmode1}, \ref{Gmode11}, \ref{Gmode111} and \ref{Gmode1111}. Further, we write the field quantization of the Dirac field $\psi$ in $R\cup L$ in terms of the local modes in $R$ and $L$ as well as in terms of global modes (we have suppressed the subscript for $s=1$ from now onwards), given as follows\\

\begin{align}
\label{localmode}
    \psi(x)&=\sum_{n}\int\frac{d\omega d k_z}{2\pi } \Bigg[a_{\rm }(\omega,k_z,n)_{\mathit{H^-_R}}U_{n}(x;\omega,k_z)_{\mathit{H^-_R}}+b^{\dagger}_{\rm }(\omega,k_z,n)_{\mathit{H^-_R}}V_{n}(x;\omega,k_z)_{\mathit{H^-_R}}\nonumber\\&+a_{\rm}(\omega,k_z,n)_{\mathit{I^-_L}}U_{n}(x;\omega,k_z)_{\mathit{I^-_L}}+b^{\dagger}_{\rm }(\omega,k_z,n)_{\mathit{I^-_L}}V_{n}(x;\omega,k_z)_{\mathit{I^-_L}}\Bigg] \nonumber\\
&=\sum_{n}\int\frac{d\omega d k_z}{2\pi } \Bigg[a_{\rm }(\omega,k_z,n)_{\mathit{I^+_R}}U_{n}(x;\omega,k_z)_{\mathit{I^+_R}}+b^{\dagger}_{\rm }(\omega,k_z,n)_{\mathit{I^+_R}}V_{n}(x;\omega,k_z)_{\mathit{I^+_R}} \nonumber\\ &+ a_{\rm}(\omega,k_z,n)_{\mathit{H^+_L}}U_{n}(x;\omega,k_z)_{\mathit{H^+_L}}+b^{\dagger}_{\rm}(\omega,k_z,n)_{\mathit{H^+_L}}V_{n}(x;\omega,k_z)_{\mathcal{H^+_L}}\Bigg]
   \end{align}

in terms of local modes,
whereas in terms of global modes, it is as follows
\begin{equation}
\label{glabalfield}
    \psi(x)=\sum_{n}\int\frac{d\omega d k_z}{2\pi } \Bigg[c_{\rm 1}(\omega,k_z,n)\phi_1^G(x)+d^{\dagger}_{\rm 1}(\omega,k_z,n)\phi_2^G(x)+c_{\rm 2}(\omega,k_z,n)\phi_4^{G}(x)+d^{\dagger}_{\rm 2}(\omega,k_z,n)\phi_3^{G}(x)\Bigg]
\end{equation}
Comparing \ref{glabalfield} and \ref{localmode}, we obtain the Bogoliubov relations,
\begin{equation}
\label{Bogoglobal}
    \begin{split}
    c_1=\frac{1}{\sqrt{2 \cosh \frac{\omega \pi}{a}}}\big(e^{\frac{\pi \omega}{2a}}a_{\mathit{H^-_{R}}}\left(\omega,k_z,n\right)-e^{-\frac{\pi \omega}{2a}}b^{\dagger}_{\mathit{I^-_{L}}}\left(-\omega,-k_z,n\right)\big),\\ d_1^{ \dagger}=\frac{1}{\sqrt{2 \cosh \frac{\omega \pi}{a}}}\big(e^{\frac{\pi \omega}{2a}}a_{\mathit{I^-_{L}}}(\omega,k_z,n)-e^{-\frac{\pi \omega}{2a}} b^{\dagger}_{\mathit{H^-_{R}}}(-\omega,-k_z,n)\big)
    \end{split}
\end{equation}
Now, using \ref{Bogoin_out} and \ref{Bogoglobal}
we can find out the relation between global and local $\mathit{out}$ operators which gives
\begin{equation}
    \label{Global_outBC}
    c_1=\frac{1}{\sqrt{2 \cosh\frac{\omega \pi}{a}}}\Bigg(e^{\frac{\pi \omega}{2a}}\alpha_1a_{\mathit{I^+_{R}}}-e^{\frac{\pi \omega}{2a}}\beta_1^*b^{\dagger}_{\mathit{I^+_{R}}}-e^{-\frac{\pi \omega}{2a}}\alpha_1^*b^{\dagger}_{\mathit{H^+_L}}+e^{-\frac{\pi \omega}{2a}}\beta_1a_{\mathit{H^+_L}}\Bigg)
\end{equation}
Similarly, there will be another set of creation and annihilation operator $\left(c_2, d_2^{\dagger}\right)$ corresponding to other set of global mode. The global vacuum can therefore be defined as $|0\rangle=|0\rangle^{1} \otimes |0\rangle^{2} $, where $|0\rangle^{1}$ is annihilated by $\left(c_1, d_1\right)$ and $|0\rangle^{2}$ is annihilated by $\left(c_2, d_2\right)$. We will work with only $|0\rangle^{1}$ only as the other will have similar results. Using the Bogoliubov relationship we can write $|0\rangle^{1}$ in terms of the local $\mathit{out}$ $R-L$ vacuum. We are now ready to compute the number density and entanglement.  
\section{The number density and logarithmic negativity}\label{S4}
\subsection{The number density of the global vacuum}
The local $\mathit{in}$ vacuum $\lvert 0\rangle_R^{\mathit{H_{R}^{-}}}$, $\lvert 0\rangle_L^{\mathit{I_{L}^{-}}}$ are defined as,\\
\begin{equation}
    \label{in vaccum}
    a_{\mathit{H^-_{R}}}\lvert 0\rangle^{\mathit{H^-_R}}=b_{\mathit{H^-_{R}}}\lvert0\rangle^{\mathit{H^-_R}}=0 \;,\;\;a_{\mathit{I^-_{L}}}\lvert0\rangle^{\mathit{I^-_L}}=b_{\mathit{I^-_{L}}}\lvert0\rangle^{\mathit{I^-_L}}=0
\end{equation}
and local $\mathit{out}$ vacuum $|0\rangle^{\mathit{I^+_R}}$, $|0\rangle^{\mathit{H^+_L}}$ as,
\begin{equation}
    \label{out vaccum}
    a_{\mathit{I^+_{R}}}\lvert0\rangle^{\mathit{I^+_R}}=b_{\mathit{I^+_{R}}}\lvert 0\rangle^{\mathit{I^+_R}}=0,\;\;a_{\mathit{H^+_{L}}}\lvert 0\rangle^{\mathit{H^+_L}}=b_{\mathit{H^+_{L}}}\lvert0\rangle^{\mathit{H^+_L}}=0
\end{equation}
The $\mathit{in}$ and $\mathit{out}$ vacuum on right $R$ wedge are related by
\begin{equation}
   \label{in_out_R}
    \lvert 0 \rangle_{\mathit{H^-_R}} = \alpha_1 \lvert 0_k 0_{-k} \rangle_{\mathit{I^+_R}}+\beta_1 \lvert 1_k 1_{-k} \rangle_{\mathit{I^+_R}} 
\end{equation}
The global vacuum is defined as
\begin{equation}
    \label{globalvac1}
    c_\sigma|0\rangle^\sigma = d_\sigma|0\rangle^\sigma=0
\end{equation}
where $\sigma=1,2$. From the Bogoliubov relation of the preceding section \ref{Global_outBC}, we express global $\mathit{in}$ vacuum in terms of local $\mathit{out}$ vacuum as follows
\begin{equation}
\label{global_vacua}
\begin{split}
    |0\rangle^{1} \equiv|0_k0_{-k}\rangle^1 =\frac{1}{(1+e^{-\frac{2\pi \omega}{a}})^{\frac{1}{2}}} \bigg(\alpha_1^{2}|0_k 0_{-k};0_k 0_{-k}\rangle^{\mathit{I^+_R};\mathit{H^+_L}}+\beta_1^{* 2}|1_k 1_{-k};1_k 1_{-k}\rangle^{\mathit{I^+_R};\mathit{H^+_L}}\\+\alpha_1 \beta_1^{*}\big(|1_k 1_{-k};0_k 0_{-k}\rangle^{\mathit{
    I^+_R};\mathit{H^+_L}}+|0_k 0_{-k};1_k 1_{-k}\rangle^{\mathit{I^+_R};\mathit{H^+_L}}\big)+e^{-\frac{\pi \omega}{a}}|1_k 0_{-k};0_k 1_{-k}\rangle^{\mathit{I^+_R};\mathit{H^+_L}}\bigg)
    \end{split}
\end{equation}
 here the first two entries corresponds to $R$, whereas the last two corresponds to $L$ and $^1\langle 0|0\rangle^{1}=1$. The Hilbert space $\mathcal{H}$ is constructed by the tensor product, $\mathcal{H} =  \mathcal{H}_k^\mathit{R} \otimes \mathcal{H}_{-k}^\mathit{R}\otimes \mathcal{H}_k^\mathit{L} \otimes \mathcal{H}_{-k}^\mathit{L}$, where $\mathcal{H}_{k}^\mathit{R}$ ($\mathcal{H}_k^\mathit{L}$) and $\mathcal{H}_{-k}^\mathit{R}$ ($\mathcal{H}_{-k}^\mathit{L}$) are the Hilbert spaces of the modes of the particle and the antiparticle, respectively and the superscript $R$ and $L$ corresponds to right and left wedge respectively. We find the spectra of pair creation in terms of the number density $(N)$ 
 \begin{equation}
     \label{vaccum_ND}
    N=^1\langle0|\;a_{\mathit{I^+_{R}}}^\dagger a_{\mathit{I^+_{R}}}|0\rangle^1=\frac{ \lvert\beta_1\rvert^2 e^{\frac{2 \pi \omega}{a}} }{1+e^{\frac{2\pi \omega}{a}}}+\frac{1}{1+e^{\frac{2\pi \omega}{a}}}
 \end{equation}
 \begin{figure}
     \begin{center}
     \includegraphics[scale=.53]{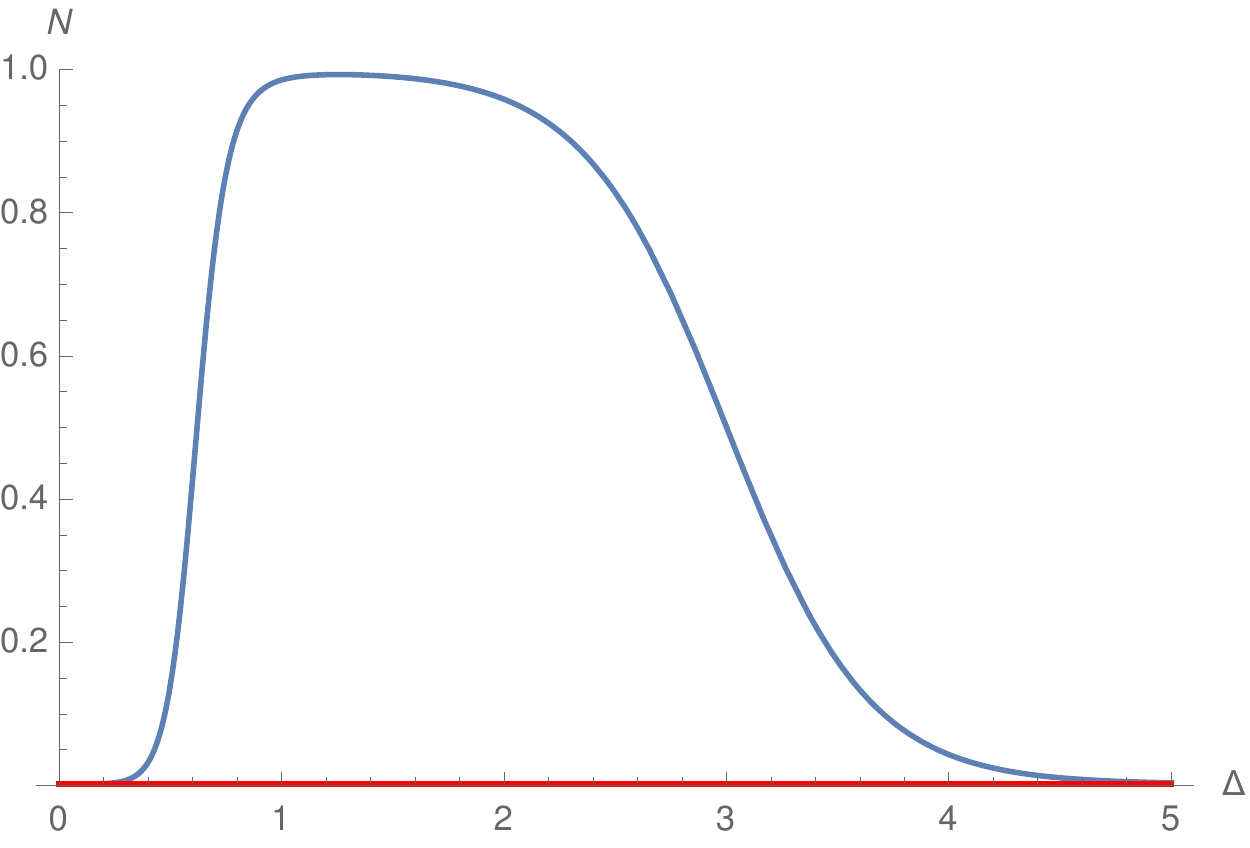}
     \caption{Number density with respect to the global vacuum, $|0\rangle^1$. As we have discussed in main text we have plotted \ref{vaccum_ND} (blue curve) and \ref{Planck_spectrum} (red curve) variation with respect to the parameter $\Delta= \frac{m^2 + (2 n + 1) q B}{q E}$, where we have taken $\frac{\omega}{a}=1$. For a given mode, the number density first monotonically increases with increase in $\Delta$ then reaches to a plateau, after that decreases monotonically with increasing $\Delta$.}
     \label{fig:ND}
     \end{center}
 \end{figure}
 here $\lvert\beta_1\rvert^2$ is given by \ref{beta2} in terms of variable $\Delta$ defined in \ref{ell}. In \ref{vaccum_ND}, the first term on the right-hand side depends on the parameter $\Delta$ and $a$, whereas the second term is independent of parameter $\Delta$. 
 For $\Delta \to \infty$ (i.e., $E \to 0$ or $q \to 0$ or $B \to \infty$), \ref{vaccum_ND} reduces to
 \begin{equation}
     \label{Planck_spectrum}
 N = \frac{1}{1+e^{\frac{2 \pi \omega}{a}}}
\end{equation}
The result \ref{Planck_spectrum} is the fermionic Planck spectrum with temperature $T=a/2\pi$, which is the usual Unruh temperature observed by an observer moving with uniform acceleration \cite{ SHN:2020, roy} in the Minkowski spacetime. Similar fermionic spectra is obtained in \cite{SSSS} for zero electric field in de Sitter spacetime; there, the non-zero number density was due to the gravitational field. \\
In the limit of vanishing acceleration $a=0$ the number density given by \ref{vaccum_ND} vanishes. Perhaps it is surprising that one would expect to reproduce the Schwinger effect in the Minkowski spacetime within this limit. This apparent ambiguity is due to the fact that the quantization of a charged field in the Rindler coordinates differs from the Minkowski coordinates. It leads to an unequal distribution of particles and antiparticles in the particular Rindler wedge in the background electric field, known as charge polarisation. Although, the total charge in the Minkowski spacetime is always conserved (i.e., the sum of charges in both the wedges). It was pointed out in \cite{Rindler1+1} for the charged scalar field in the presence of a constant strength background electric field in the $(1+1)$-Rindler spacetime. \ref{vaccum_ND} represents the number density of particles near $I^{+}_R$ region of the right Rindler wedge for the Minkowski vacuum; it vanishes in the limit $a=0$. One might expect the non-zero number density of antiparticles and particles near $I^{+}_R$ and $I^{+}_L$ regions, respectively. A similar analysis has been done for different regions of the particular Rindler wedge for the charged scalar field; for details, the reader can refer to \cite{Rindler1+1}. Note, at the limit of vanishing acceleration, the number density of particles near region $I^+_R$ given by \ref{beta2} also vanishes.

We have plotted number density \ref{vaccum_ND} vs $\Delta$ in \ref{fig:ND} (blue curve) and it is non-monotonic unlike the case of inertial observer. The number density $(N)$ increases monotonically with an increase in $\Delta$ and then reaches a plateau near $\Delta \approx 1$; after that, it monotonically decreases with an increase in $\Delta$. However, for $\Delta \to \infty$, 
number density have some finite value (i.e. $N \approx 0.06$) which corresponds to \ref{Planck_spectrum} (shown by red curve in \ref{fig:ND}). This non-vanishing number density at $\Delta \to \infty$ is due to the non-zero acceleration of uniformly accelerated observer and also the expected behaviour of the Dirac field due to the form of the fermionic spectra \cite{Dirac2, Fuentes:2010dt}. 
\subsection{The logarithmic negativity}
Next, we wish to compute entanglement between the particles and antiparticles in the $R$ and $L$ regions, respectively. The state, which represents the particle-antiparticle of the $R$ and $L$ regions, is characterized by a mixed state density matrix given by \ref{densityRL}. Logarithmic negativity is a good measure to compute entanglement for a mixed state, therefore we computed logarithmic negativity for \ref{densityRL}.

For mixed states, there is a measure of the entanglement of bipartite states~\cite{Zyczkowski:1998yd, Vidal:2002zz}, called the entanglement negativity, defined as
$
    \mathcal{N}(\rho_{AB})
=
	\frac{1}{2} \left(\lvert \lvert \rho_{AB}^{T_A} \rvert \rvert_1-1\right)
$, 
where $\rho_{AB}^{\text{T}_A}$ is the partial transpose of $\rho_{AB}$ with respect to the subspace of $A$, i.e., $\left( \lvert i\rangle_{\! A} \hspace{-0.2ex} \langle n \rvert \otimes \lvert j \rangle_{\! B} \hspace{-0.2ex} \langle \ell \rvert \right)^{\text{T}_A}: = \lvert n \rangle_{\! A} \hspace{-0.2ex} \langle i \rvert \otimes \lvert j \rangle_{\! B} \hspace{-0.2ex} \langle \ell \rvert$.
Here, $\lvert \lvert \rho_{AB}^{T_A} \rvert\rvert_{1}$ is the trace norm, $\lvert \lvert\rho_{AB}^{T_A}\rvert \rvert_{1} = \sum_{i=1}^{\text{all}} \lvert \mu_i \rvert$, where $\mu_i$ is the $i$-th eigenvalue of $\rho_{AB}^{T_A}$. 
The logarithm of $\lvert \lvert \rho_{AB}^{T_A} \rvert \rvert_1$ is called the logarithmic negativity, which can be written as $L_{N} (\rho_{AB})
=
	\log \left( 1 + 2 \mathcal{N}(\rho_{AB}) \right)$.
These quantities are entanglement monotones that do not increase under local operations and classical communications.
These quantities measure a violation of the positive partial transpose (PPT) in $\rho_{AB}$.
The PPT criterion can be stated as follows. If $\rho_{AB}$ is separable, the eigenvalues of $\rho_{AB}^{T_A}$ are non-negative.
Hence, if $\mathcal{N} \not = 0$ ($L_N \not = 0$), $\rho_{AB}$ is an entangled state.
On the other hand, if $\mathcal{N} = 0$ ($L_N = 0$), we cannot judge the existence of the entanglement from this measure since there exist PPT and entangled states in general.
However, the logarithmic negativity can be helpful since it is a calculable measure, and more discussions on it can be found in e.g.~\cite{Horodecki:2009zz}.

\begin{figure}[h]
\begin{center}
		\includegraphics[scale=.53]{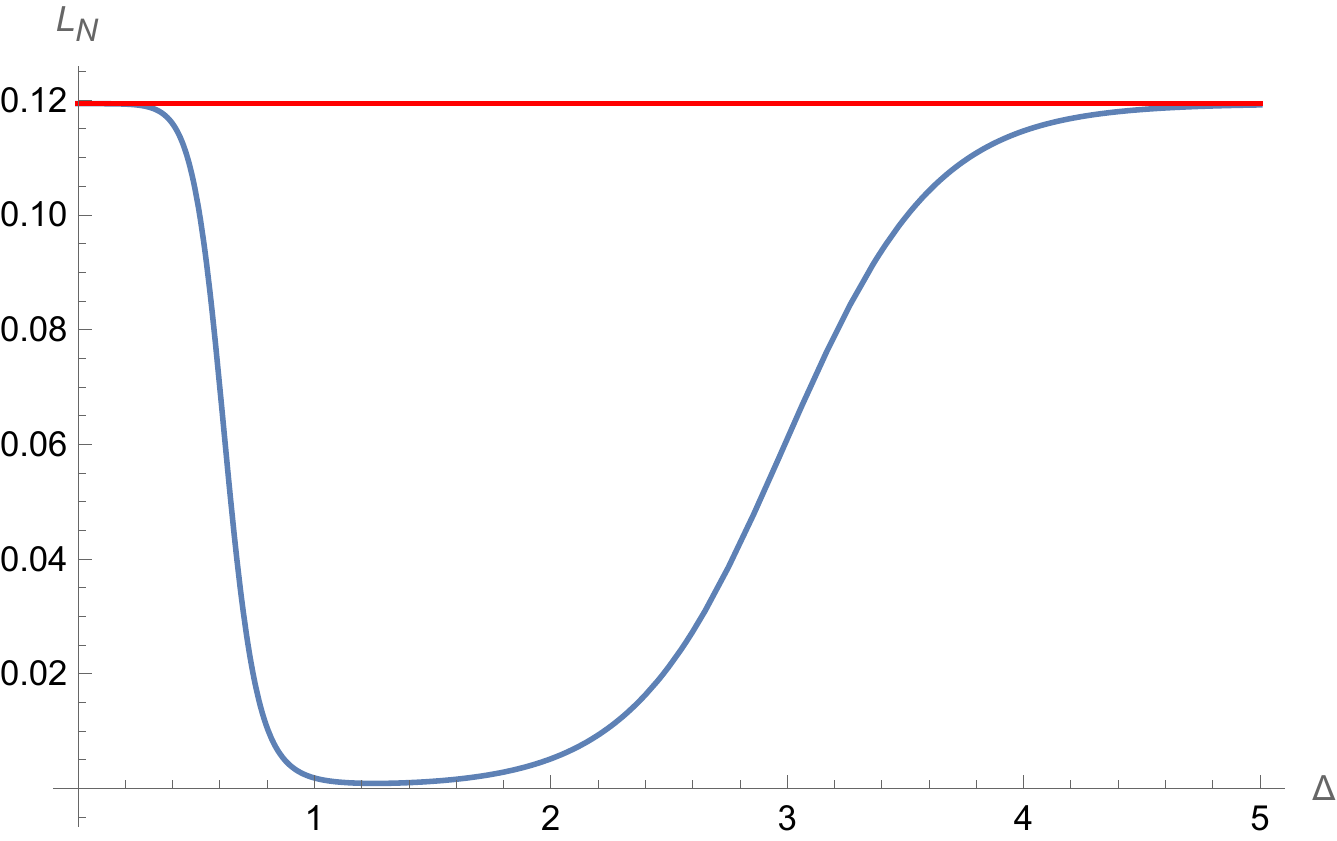}\hspace{1.0cm}
		\caption{Logarithmic negativity between the particles and antiparticles in $R$ and $L$ wedges respectively. As we have discussed in main text we have plotted \ref{LN} (blue curve) and \ref{LNE} (red curve) variation with respect to the parameter $\Delta= \frac{m^2 + (2 n + 1) q B}{q E}$, where we have taken $\frac{\omega}{a}=1$. Logarithmic negativity first monotonically decreases with increase in $\Delta$ then reaches to a plateau, after that increases monotonically with increasing $\Delta$.
		}
		\label{fig:LNforVacuum}
		\end{center}
\end{figure}

The total density operator for global vacuum is $\rho_{global}=|0\rangle^{1} ~^{1}\langle0|$. We obtain the reduced density operator for particles of $R$ wedge and antiparticles of $L$ wedge by tracing out antiparticles of $R$ wedge and particles of $L$ wedge given as
\begin{eqnarray}
 \label{densityRL}
    \rho^{p;a}_{R;L}&=& \frac{1}{1+e^{-\frac{2 \pi \omega}{a}}}\Big(\lvert\alpha_1\rvert^4 \lvert00\rangle \langle00\rvert +(\lvert\beta_1\rvert^4+e^{-\frac{2 \pi \omega}{a}})\lvert 11\rangle \langle11\rvert+\lvert\alpha_1\rvert^2 \lvert\beta_1\rvert^2(\lvert10\rangle \langle10\rvert+\lvert01\rangle \langle01\rvert) \nonumber\\
    &+&e^{-\frac{\pi \omega}{a}}(\alpha_1^2\lvert10\rangle \langle10\rvert+\alpha_1^{*2}\lvert01\rangle \langle01\rvert)\Big)
\end{eqnarray}
Now using the definition of logarithmic negativity ($L_N$) for $\rho^{p;a}_{R;L}$, we have
\begin{equation}
    \label{LN}
    L_N= \log_2\Big[1+\frac{e^{-\frac{\pi \omega}{a}}(\alpha_1^2+\alpha_1^{*2})}{1+e^{-\frac{2\pi \omega}{a}}}\Big]
\end{equation}
$L_N$ is a function of $\Delta$ and $a$, therefore the entanglement depends on the motion of the observer and the strength of electromagnetic field. For $\Delta \to \infty$,
\begin{equation}
\label{LNE}
L_N=\log_2\Big[1+\frac{2}{e^{\frac{\pi \omega}{a}}+e^{-\frac{\pi \omega}{a}}}\Big]
\end{equation}
which represents the usual $R-L$ entanglement. We have plotted \ref{LN} 
and \ref{LNE} vs. $\Delta$ in \ref{fig:LNforVacuum}. In the presence of background electromagnetic field, initially $L_N$ decreases monotonically with an increase in $\Delta$, whereas near $\Delta \approx1$, it reaches to a plateau and further increases monotonically with the increase in $\Delta$ (blue curve). Whereas the red curve represents \ref{LNE}. 
The behaviour of logarithmic negativity of $ \rho^{p; a}_{R; L}$, is  may be due to the mixed state structure of density matrix and the choice of the reduced density matrix.

\section{Summary and outlook}
\label{S5}
This work has investigated the effect of constant background electromagnetic fields on particle creation in the Rindler spacetime. Also, some aspects of quantum entanglement between the created particles is discussed very briefly. We have found the $\mathit{in}$ and $\mathit{out}$ local modes for both R and L wedges and the Bogoliubov relationship between them is derived in \ref{section : S2}. The global modes are constructed from local modes, and the Bogoliubov transformations between relevant creation and annihilation operators are obtained in \ref{S3}. Using, these results we have written the squeezed state expansion of global vacuum in terms of local $\mathit{out}$ vacuum basis. In \ref{S4}, further, we have obtained the number density of created particles with respect to the global vacuum and logarithmic negativity between
the particles and antiparticles on the R and L wedges respectively.

The main characteristic of this problem is that it involves two acceleration parameters: the acceleration of the Rindler observer $a$ and the natural acceleration of the charged quanta $(qE/m)$ due to the background electric field. Therefore, there are two sources of particle creation here: the Schwinger as well as Unruh effects. As we have discussed in \ref{introduction}, the magnetic field alone cannot be expected to create vacuum instability, but it may affect the  impact of the electric field on the same. Indeed, from \ref{beta2} which represents the number density for local vacuum, it is clear that the magnetic field holds no role in pair creation in the absence of electric field. Also, in the presence of the electric field, the magnetic field opposes the effect of the electric field. Moreover, from \ref{beta2}, it is clear that the particle creation due to the Schwinger effect depends upon the observer's motion characterized by the acceleration parameter i.e. $a$. However, in this case also if we turn off the electric field the magnetic field does not affect whatsoever the particle creation.

Further, in \ref{fig:ND}, we have taken into account the variation of number density of created particles with respect to the global vacuum \ref{vaccum_ND} with respect to the parameter $\Delta$ defined in \ref{ell}, and it comes out to be non-monotonic. Next, we found logarithmic negativity between the particles and antiparticles in R and L wedge, respectively, to gain insight into the entanglement property of created particles. In \ref{fig:LNforVacuum} we have shown the variation of \ref{LN} concerning parameter $\Delta$, 
its behaviour is non-monotonic and depends on the choice of the reduced density matrix we obtained from the full density operator, $\rho_{\text{global}}$. One can further extend this analysis with time-dependent electromagnetic fields. Also, we wish to extend this work to study the correlation between different sectors of an initially entangled state constructed by two or more fields.

\section*{Acknowledgement}
I thank Sourav Bhattacharya for his constant support and suggestions to improve the manuscript. I also thank Md Sabir Ali and Shankhadeep Chakrabortty, for the fruitful discussions and comments on the manuscript.

\appendix
\labelformat{section}{Appendix #1}
\section{Explicit form of the mode functions and normalizations}\label{A}
\begin{equation}
\begin{split}
\label{fullmode1}
    U_{1,n}(x)_{\mathit{H^-_R}}=\frac{1}{N_1\;e^{\frac{a x}{2}}}\bigg(\frac{i \epsilon_4}{e^{a x}}\partial_t -\frac{i \epsilon_4 }{e^{a x}}\partial_x-\epsilon_2\partial_2-i\epsilon_2\partial_3-\frac{q E e^{a x}}{2 a} \epsilon_4+ q B y \epsilon_2 + m \epsilon_1\bigg)\\ \times e^{-i \omega (t-x)} e^{-i k_z z} e^{a x} e^{-\frac{i q E e^{2 a x}}{4 a^2}} U(\lambda_1, \nu, \xi)H_1(y) 
    \end{split}
\end{equation}
 \begin{equation}
\begin{split}
\label{fullmode2}
    U_{1,n}(x)_{\mathit{I^-_R}}=\frac{1}{M_1\; e^{\frac{a x}{2}}}\bigg(\frac{i \epsilon_4}{e^{a x}}\partial_t -\frac{i \epsilon_4 }{e^{a x}}\partial_x-\epsilon_2\partial_2-i\epsilon_2\partial_3-\frac{q E e^{a x}}{2 a} \epsilon_4+ q B y \epsilon_2 + m \epsilon_1\bigg)\\\times e^{-i \omega (t+x)} e^{-i k_z z} e^{a x} e^{\frac{i q E e^{2 a x}}{4 a^2}} (L(-\lambda_1, \nu-1, \xi))^* H_1(y) 
    \end{split}
\end{equation}
\begin{equation}
\begin{split}
\label{fullmode3}
    U_{2,n}(x)_{\mathit{H^-_R}}=\frac{1}{N_2\;e^{\frac{a x}{2}}}\bigg(\frac{i \epsilon_3}{e^{a x}}\partial_t -\frac{i \epsilon_3 }{e^{a x}}\partial_x-\epsilon_1\partial_2-i\epsilon_1\partial_3-\frac{q E e^{a x}}{2 a} \epsilon_3+ q B y \epsilon_1 + m \epsilon_2\bigg)\\ \times e^{-i \omega (t-x)} e^{-i k_z z} e^{a x} e^{-\frac{i q E e^{2 a x}}{4 a^2}} U(\lambda_2, \nu, \xi) H_2(y)
    \end{split}
\end{equation}
\begin{equation}
\begin{split}
\label{fullmode4}
     U_{2,n}(x)_{\mathit{I^-_R}}=\frac{1}{M_2\; e^{\frac{a x}{2}}}\bigg(\frac{i \epsilon_3}{e^{a x}}\partial_t -\frac{i \epsilon_3 }{e^{a x}}\partial_x-\epsilon_1\partial_2-i\epsilon_1\partial_3-\frac{q E e^{a x}}{2 a} \epsilon_3+ q B y \epsilon_1 + m \epsilon_2\bigg)\\ \times e^{-i \omega (t+x)} e^{-i k_z z} e^{a x} e^{\frac{i q E e^{2 a x}}{4 a^2}} (L(-\lambda_2, \nu-1, \xi))^* H_2(y) 
    \end{split}
\end{equation}
\begin{equation}
\begin{split}
\label{fullmode1a}
   V_{1,n}(x)_{\mathit{H^-_R}}=\frac{1}{P_1\;e^{\frac{a x}{2}}}\bigg(-\frac{i \epsilon_2}{e^{a x}}\partial_t +\frac{i \epsilon_2 }{e^{a x}}\partial_x-\epsilon_4\partial_2-i\epsilon_4\partial_3+\frac{q E e^{a x}}{2 a} \epsilon_2+ q B y \epsilon_4 + m \epsilon_3\bigg)\\ \times e^{i \omega (t-x)} e^{i k_z z} e^{a x} e^{\frac{i q E e^{2 a x}}{4 a^2}} e^{\xi}U(\nu-\lambda_1,\nu,\xi)H_1(y_-)
    \end{split}
\end{equation}
\begin{equation}
\begin{split}
\label{fullmode2a}
   V_{1,n}(x)_{\mathit{I^-_R}}=\frac{1}{R_1\;e^{\frac{a \xi}{2}}}\bigg(-\frac{i \epsilon_2}{e^{a x}}\partial_t +\frac{i \epsilon_2 }{e^{a x}}\partial_x-\epsilon_4\partial_2-i\epsilon_4\partial_3+\frac{q E e^{a x}}{2 a} \epsilon_2+ q B y \epsilon_4 + m \epsilon_3\bigg)\\ \times e^{i \omega (t+x)} e^{i k_z z} e^{a x} e^{-\frac{i q E e^{2 a x}}{4 a^2}}(\xi^{1-\nu} L(\nu-\lambda_s-1,1-\nu,\xi))^* H_1(y_-)
    \end{split}
\end{equation}
\begin{equation}
\begin{split}
\label{fullmode3a}
   V_{2,n}(x)_{\mathit{H^-_R}}=\frac{1}{P_2\;e^{\frac{a x}{2}}}\bigg(-\frac{i \epsilon_1}{e^{a x}}\partial_t +\frac{i \epsilon_1 }{e^{a x}}\partial_x-\epsilon_3\partial_2-i\epsilon_3\partial_3+\frac{q E e^{a x}}{2 a} \epsilon_1+ q B y \epsilon_3 + m \epsilon_4\bigg)\\ \times e^{i \omega (t-x)} e^{i k_z z} e^{a x} e^{\frac{i q E e^{2 a x}}{4 a^2}}e^{\xi}U(\nu-\lambda_2,\nu,\xi)H_2(y_-)
    \end{split}
\end{equation}
\begin{equation}
\begin{split}
\label{fullmode4a}
    V_{2,n}(x)_{\mathit{I^-_R}}=\frac{1}{R_2\;e^{\frac{a x}{2}}}\bigg(-\frac{i \epsilon_1}{e^{a x}}\partial_t +\frac{i \epsilon_1 }{e^{a x}}\partial_x-\epsilon_3\partial_2-i\epsilon_3\partial_3+\frac{q E e^{a x}}{2 a} \epsilon_1+ q B y \epsilon_3 + m \epsilon_4\bigg)\\ \times e^{i \omega (t+x)} e^{i k_z z} e^{a x} e^{-\frac{i q E e^{2 a x}}{4 a^2}} (\xi^{1-\nu}L(\nu-\lambda_2-1,1-\nu,\xi))^* H_2(y_-)
    \end{split}
\end{equation} 
here in \ref{fullmode1a}, \ref{fullmode2a}, \ref{fullmode3a} and \ref{fullmode4a} $y_-=\left(\sqrt{q B}y-\frac{k_z}{\sqrt{q B}}\right)$ and $H_1(y_-)= H_2(y_-)= \left(\frac{\sqrt{q B}}{2^{n+1}\sqrt{\pi}(n+1)!}\right)^{1/2}\\
e^{-y_-^2/2}\mathit{H}_{n}(y_-).$ The normalisation constants, $N_1,\,N_2,\,M_1$ and $M_2$ are given by previous section. We shall explicitly evaluate $N_1$ below, for which we choose constant time hypersurface with normal vector $n_\mu=e^{-a x} t^{\mu}$ 
\begin{equation*}
\begin{split}
   \int_{\vec{x}}  n_\mu \sqrt{|g|} \gamma^{(0)} \gamma^0 (U_{s,n}(x)_{\mathit{H^-_R}})^{\dagger} U_{s,n^\prime}(x)_{\mathit{H^-_R}}=\frac{1}{|N_s|^2} a^3 \int \bigg[\bigg(\frac{\omega}{e^{a x}}-\frac{q E}{2 a}-\frac{i}{e^{a x}}\big(a-i\frac{q e^{2 a x}}{2 a}+i \omega-2 a \lambda_1\big)\bigg)\\ \times\bigg(\frac{\omega^{\prime}}{e^{a x}}-\frac{q E}{2 a}+\frac{i}{e^{a x}}\big(a+i\frac{q e^{2 a x}}{2 a}-i \omega^{\prime}-2 a \lambda_1^*\big)\bigg) H_1^{\prime}(y) H_1(y) + (\partial_y + y_{+}\sqrt{q B})H_1^{\prime}(y)(\partial_y + y_{+}\sqrt{q B})H_1(y)+m^2\bigg]\\ \times e^{i(\omega-\omega^{\prime})(t-x)} e^{i(k_z-k_z^{\prime})z}\bigg(\frac{2 a^2}{i q E}\bigg)^{\lambda}\bigg(\frac{2 a^2}{-i q E}\bigg)^{\lambda^*}e^{-2 a x}\\
  =\frac{1}{|N_s|^2} \bigg(\frac{a^2}{ E^2}e^{-\pi\frac{m^2+S_s}{4 q E}}\bigg) \delta_{nn'} \delta(\omega-\omega') \delta(k_z-k_z')
   \end{split}
\end{equation*}
Here we have used the asymptotic limit of $U(\lambda, \nu, \xi)$  function at $x \to \infty\; (|\xi| \to \infty)$, i.e. $U(\lambda, \nu, \xi) \approx \xi^{-\lambda}$, the variable $\xi$ is defined in \ref{ell}. Note that the normalization of $U_{s,n}(x)_{\mathit{I^-_R}}$ can be done in the same way as of $U_{s,n}(x)_{\mathit{H^-_R}}$ for which we have used the asymptotic form of $L(\lambda, \nu, \xi)$ at $x \to -\infty\; (|\xi| \to 0) $, i.e. $L(\lambda, \nu, \xi) = \frac{\Gamma(\nu+\lambda+1)}{\Gamma(\lambda+1) \Gamma (\nu+1)}$ which gives $M_s$, similarly we normalize all other in modes. Set of out modes are given as follows
\begin{equation}
\begin{split}
\label{fullmode1out}
    U_{1,n}(x)_{\mathit{H^+_R}}=\frac{1}{N_1\;e^{\frac{a x}{2}}}\bigg(-\frac{i \epsilon_4}{e^{a x}}\partial_t +\frac{i \epsilon_4 }{e^{a x}}\partial_x-\epsilon_2\partial_2+i\epsilon_2\partial_3-\frac{q E e^{a x}}{2 a} \epsilon_4+ q B y \epsilon_2 + m \epsilon_1\bigg)\\ \times e^{-i \omega (t+x)} e^{-i k_z z} e^{a x} e^{\frac{i q E e^{2 a x}}{4 a^2}} (U(\lambda_1, \nu, \xi))^*H_1(y) 
    \end{split}
\end{equation}
 \begin{equation}
\begin{split}
\label{fullmode2out}
    U_{1,n}(x)_{\mathit{I^+_R}}=\frac{1}{M_1\; e^{\frac{a x}{2}}}\bigg(-\frac{i \epsilon_4}{e^{a x}}\partial_t +\frac{i \epsilon_4 }{e^{a x}}\partial_x-\epsilon_2\partial_2+i\epsilon_2\partial_3-\frac{q E e^{a x}}{2 a} \epsilon_4+ q B y \epsilon_2 + m \epsilon_1\bigg)\\ \times e^{-i \omega (t-x)} e^{-i k_z z} e^{a x} e^{-\frac{i q E e^{2 a x}}{4 a^2}}L(-\lambda_1, \nu-1, \xi) H_1(y)
    \end{split}
\end{equation}
\begin{equation}
\begin{split}
\label{fullmode3out}
    U_{2,n}(x)_{\mathit{H^+_R}}=\frac{1}{N_2\;e^{\frac{a x}{2}}}\bigg(-\frac{i \epsilon_3}{e^{a x}}\partial_t +\frac{i \epsilon_3 }{e^{a x}}\partial_x-\epsilon_1\partial_2+i\epsilon_1\partial_3-\frac{q E e^{a x}}{2 a} \epsilon_3+ q B y \epsilon_1 + m \epsilon_2\bigg)\\ \times e^{-i \omega (t+x)} e^{-i k_z z} e^{a x} e^{\frac{i q E e^{2 a x}}{4 a^2}} (U(\lambda_2, \nu, \xi))^* H_2(y)
    \end{split}
\end{equation}
\begin{equation}
\begin{split}
\label{fullmode4out}
     U_{2,n}(x)_{\mathit{I^+_R}}=\frac{1}{M_2\; e^{\frac{a x}{2}}}\bigg(-\frac{i \epsilon_3}{e^{a x}}\partial_t+\frac{i \epsilon_3 }{e^{a x}}\partial_x-\epsilon_1\partial_2+i\epsilon_1\partial_3-\frac{q E e^{a x}}{2 a} \epsilon_3+ q B y \epsilon_1 + m \epsilon_2\bigg)\\ \times e^{-i \omega (t-x)} e^{-i k_z z} e^{a x} e^{-\frac{i q E e^{2 a x}}{4 a^2}}L(-\lambda_2,\nu-1,\xi)  H_2(y) 
    \end{split}
\end{equation}
\begin{equation}
\begin{split}
\label{fullmode1aout}
   V_{1,n}(x)_{\mathit{H^+_R}}=\frac{1}{P_1\;e^{\frac{a x}{2}}}\bigg(\frac{i \epsilon_2}{e^{a x}}\partial_t -\frac{i \epsilon_2 }{e^{a x}}\partial_x-\epsilon_4\partial_2+i\epsilon_4\partial_3+\frac{q E e^{a x}}{2 a} \epsilon_2+ q B y \epsilon_4 + m \epsilon_3\bigg)\\ \times e^{i \omega (t+x)} e^{i k_z z} e^{a x} e^{\frac{-i q E e^{2 a x}}{4 a^2}} (e^{\xi}U(\nu-\lambda_1,\nu,\xi))^*H_1(y_-)
    \end{split}
\end{equation}
\begin{equation}
\begin{split}
\label{fullmode2aout}
   V_{1,n}(x)_{\mathit{I^+_R}}=\frac{1}{R_1\;e^{\frac{a x}{2}}}\bigg(\frac{i \epsilon_2}{e^{a x}}\partial_t -\frac{i \epsilon_2 }{e^{a x}}\partial_x-\epsilon_4\partial_2+i\epsilon_4\partial_3+\frac{q E e^{a x}}{2 a} \epsilon_2+ q B y \epsilon_4 + m \epsilon_3\bigg)\\ \times e^{i \omega (t-x)} e^{i k_z z} e^{a x} e^{\frac{i q E e^{2 a x}}{4 a^2}}\xi^{1-\nu}L(\nu-\lambda_1-1,1-\nu,\xi) H_1(y_-)
    \end{split}
\end{equation}
\begin{equation}
\begin{split}
\label{fullmode3aout}
   V_{2,n}(x)_{\mathit{H^+_R}}=\frac{1}{P_2\;e^{\frac{a x}{2}}}\bigg(\frac{i \epsilon_1}{e^{a x}}\partial_t -\frac{i \epsilon_1 }{e^{a x}}\partial_x-\epsilon_3\partial_2+i\epsilon_3\partial_3+\frac{q E e^{a x}}{2 a} \epsilon_1+ q B y \epsilon_3 + m \epsilon_4\bigg)\\ \times e^{i \omega (t+x)} e^{i k_z z} e^{a x} e^{\frac{-i q E e^{2 a x}}{4 a^2}} (e^{\xi} U(\nu-\lambda_2,\nu,\xi))^*H_2(y_-)
    \end{split}
\end{equation}
\begin{equation}
\begin{split}
\label{fullmode4aout}
    V_{2,n}(x)_{\mathit{I^+_R}}=\frac{1}{R_2\;e^{\frac{a x}{2}}}\bigg(\frac{i \epsilon_1}{e^{a x}}\partial_t -\frac{i \epsilon_1 }{e^{a x}}\partial_x-\epsilon_3\partial_2+i\epsilon_3\partial_3+\frac{q E e^{a x}}{2 a} \epsilon_1+ q B y \epsilon_3 + m \epsilon_4\bigg)\\ \times e^{i \omega (t-x)} e^{i k_z z} e^{a x} e^{\frac{i q E e^{2 a x}}{4 a^2}} \xi^{1-\nu} L(\nu-\lambda_2-1,1-\nu,\xi) H_2(y_-)
    \end{split}
\end{equation}

Remaining modes for $L$ region are as follows
\begin{equation}
\begin{split}
\label{fullmode1LL}
    U_{1,n}(x_L)_{\mathit{H^+_L}}=\frac{1}{N_1\;e^{\frac{a x_L}{2}}}\bigg(\frac{i \epsilon_4}{e^{a x_L}}\partial_{t_{L}} -\frac{i \epsilon_4 }{e^{a x_L}}\partial_{x_{L}}-\epsilon_2\partial_2-i\epsilon_2\partial_3-\frac{q E e^{a x_L}}{2 a} \epsilon_4+ q B y \epsilon_2 + m \epsilon_1\bigg)\\ \times e^{-i \omega (t_L-x_L)} e^{i k_z z} e^{a x_L} e^{-\frac{i q E e^{2 a x_L}}{4 a^2}} U(\lambda_1, \nu, \xi_L)H_1(y) 
    \end{split}
\end{equation}
 \begin{equation}
\begin{split}
\label{fullmode2LL}
    U_{1,n}(x_L)_{\mathit{I^+_L}}=\frac{1}{M_1\; e^{\frac{a x_L}{2}}}\bigg(\frac{i \epsilon_4}{e^{a x_L}}\partial_{t_{L}} -\frac{i \epsilon_4 }{e^{a x_L}}\partial_{x_{L}}-\epsilon_2\partial_2-i\epsilon_2\partial_3-\frac{q E e^{a x_L}}{2 a} \epsilon_4+ q B y \epsilon_2 + m \epsilon_1\bigg)\\ \times e^{-i \omega (t_L+x_L)} e^{i k_z z} e^{a x_L} e^{\frac{i q E e^{2 a x_L}}{4 a^2}}  (L(-\lambda_1, \nu-1, \xi_L))^* H_1(y) 
    \end{split}
\end{equation}
\begin{equation}
\begin{split}
\label{fullmode3L}
    U_{2,n}(x_L)_{\mathit{H^+_L}}=\frac{1}{N_2\;e^{\frac{a x_L}{2}}}\bigg(\frac{i \epsilon_3}{e^{a x_L}}\partial_{t_{L}} -\frac{i \epsilon_3 }{e^{a x_L}}\partial_{x_{L}}-\epsilon_1\partial_2-i\epsilon_1\partial_3-\frac{q E e^{a x_L}}{2 a} \epsilon_3+ q B y \epsilon_1 + m \epsilon_2\bigg)\\ \times e^{-i \omega (t_L-x_L)} e^{i k_z z} e^{a x_L} e^{-\frac{i q E e^{2 a x_L}}{4 a^2}} U(\lambda_2, \nu, \xi_L) H_2(y)
    \end{split}
\end{equation}
\begin{equation}
\begin{split}
\label{fullmode4L}
     U_{2,n}(x_L)_{\mathit{I^+_L}}=\frac{1}{M_1\; e^{\frac{a x_L}{2}}}\bigg(\frac{i \epsilon_3}{e^{a x_L}}\partial_{t_{L}} -\frac{i \epsilon_3 }{e^{a x_L}}\partial_{x_{L}}-\epsilon_1\partial_2-i\epsilon_1\partial_3-\frac{q E e^{a x_L}}{2 a} \epsilon_3+ q B y \epsilon_1 + m \epsilon_2\bigg)\\ \times e^{-i \omega (t_L+x_L)} e^{i k_z z} e^{a x_L} e^{\frac{i q E e^{2 a x_L}}{4 a^2}}  (L(-\lambda_1, \nu-1, \xi))^* H_2(y) 
    \end{split}
\end{equation}

\begin{equation}
\begin{split}
\label{fullmode1aLL}
   V_{1,n}(x_L)_{\mathit{H^+_L}}=\frac{1}{P_1\;e^{\frac{a x_L}{2}}}\bigg(-\frac{i \epsilon_2}{e^{a x_L}}\partial_{t_L} +\frac{i \epsilon_2 }{e^{a x}}\partial_x-\epsilon_4\partial_2-i\epsilon_4\partial_3+\frac{q E e^{a x_L}}{2 a} \epsilon_2+ q B y \epsilon_4 + m \epsilon_3\bigg)\\ \times e^{i \omega (t_L-x_L)} e^{-i k_z z} e^{a x_L} e^{\frac{i q E e^{2 a x_L}}{4 a^2}} e^{\xi}U(\nu-\lambda_1,\nu,\xi)H_1(y_-)
    \end{split}
\end{equation}
\begin{equation}
\begin{split}
\label{fullmode2aLL}
   V_{1,n}(x_{L})_{\mathit{I^+_L}}=\frac{1}{R_1 \, e^{\frac{a x_L}{2}}}\bigg(-\frac{i \epsilon_2}{e^{a x_L}}\partial_{t_L} +\frac{i \epsilon_2 }{e^{a x_L}}\partial_{x_{L}}-\epsilon_4\partial_2-i\epsilon_4\partial_3+\frac{q E e^{a x_L}}{2 a} \epsilon_2+ q B y \epsilon_4 + m \epsilon_3\bigg)\\
   \times e^{i \omega (t_{L}+x_{L})} e^{-i k_{z} z} e^{a x_{L}} e^{-\frac{i q E e^{2 a x_{L}}}{4 a^2}}(\xi_{L}^{1-\nu} L(\nu-\lambda_{1}-1,1-\nu,\xi))^{*} H_{1} (y_{-})
    \end{split}
\end{equation}
\begin{equation}
\begin{split}
\label{fullmode3aL}
   V_{2,n}(x_L)_{\mathit{H^+_L}}=\frac{1}{P_2\;e^{\frac{a x_L}{2}}}\bigg(-\frac{i \epsilon_1}{e^{a x_L}}\partial_{t_L} +\frac{i \epsilon_1 }{e^{a x_L}}\partial_{x_L}-\epsilon_3\partial_2-i\epsilon_3\partial_3+\frac{q E e^{a x_L}}{2 a} \epsilon_1+ q B y \epsilon_3 + m \epsilon_4\bigg)\\ \times e^{i \omega (t_L-x_L)} e^{-i k_z z} e^{a x} e^{\frac{i q E e^{2 a x_L}}{4 a^2}} e^{\xi}U(\nu-\lambda_2,\nu,\xi)H_2(y_-)
    \end{split}
\end{equation}
\begin{equation}
\begin{split}
\label{fullmode4aL}
    V_{2,n}(x_L)_{\mathit{I^+_L}}=\frac{1}{R_2\;e^{\frac{a x_L}{2}}}\bigg(-\frac{i \epsilon_1}{e^{a x_L}}\partial_{t_L} +\frac{i \epsilon_1 }{e^{a x_L}}\partial_{x_L}-\epsilon_3\partial_2-i\epsilon_3\partial_3+\frac{q E e^{a x_L}}{2 a} \epsilon_1+ q B y \epsilon_3 + m \epsilon_4\bigg)\\ \times  e^{i \omega (t_L+x_L)} e^{-i k_z z} e^{a x_L} e^{-\frac{i q E e^{2 a x_L}}{4 a^2}} (\xi_L^{1-\nu} L(\nu-\lambda_2-1,1-\nu,\xi_L))^*  H_2(y_-)
    \end{split}
\end{equation} 
where the normalization constants are \\\begin{center}
    $N_s=e^{-\frac{\pi \Delta}{2}} \cosh{\frac{\pi \omega}{a}} \Big(\frac{\sinh{\pi \Delta} \cosh{\pi (\Delta - \frac{\omega}{a})}}{\cosh^3{\pi(\Delta-\frac{\omega}{a})}+e^{-\pi \Delta} \sinh^3{\pi \Delta}}\Big)^{\frac{1}{2}} $, $ M_s=e^{-\frac{\pi \Delta}{2}}  \sqrt{\frac{\pi}{\Delta}}$, $P_s= \Big(\frac{6n+1}{eB}+m^2\Big)^{\frac{1}{2}}$  and $R_s =\sqrt{\pi}$.
\end{center}
here $s=1,2$.
\section{Limits of modes near null infinities}\label{B}
\begin{equation}
\label{fullmode1IRlim}
    U_{s,n}(x)_{\mathit{H^-_R}}=\frac{1}{N_s\;e^{\frac{a x}{2}}}\bigg(ie^\mu_a \gamma^{(a)}\partial_\mu-q A_\mu e^\mu_a \gamma^{(a)}+m\bigg)e^{-i \omega (t-x)} e^{-i k_z z} e^{-a x} e^{-\frac{i q E e^{2 a x}}{4 a^2}} \Big(-\frac{i q E}{2a^2}\Big)^{-\lambda_s}e^{-\frac{ia (m^2+S_s)x}{qE}}H_s(y)\epsilon_s
\end{equation}
\begin{equation}
\label{fullmode1aIRlim1}
    V_{s,n}(x)_{\mathit{H^-_R}}=\frac{1}{P_s e^{\frac{a x}{2}}} \bigg(ie^\mu_a \gamma^{(a)}\partial_\mu-q A_\mu e^\mu_a \gamma^{(a)}+m\bigg)e^{i \omega (t+x)} e^{i k_z z} e^{-a x} e^{-\frac{i q E e^{2 a x}}{4 a^2}} \Big(-\frac{i q E}{2a^2}\Big)^{-\nu^*+\lambda_s}e^{\frac{i a (m^2+S_s) x}{qE}} H_s(y) \epsilon_s,
\end{equation}
\begin{equation}
\label{fullmode1ILlim}
    U_{s,n}(x)_{\mathit{I^-_L}}=\frac{1}{M_s\;e^{\frac{a x_L}{2}}}\bigg(ie^\mu_a \gamma^{(a)}\partial_\mu-q A_\mu e^\mu_a \gamma^{(a)}+m\bigg)e^{-i \omega (t_L+x_L)} e^{i k_z z} e^{a x_L} e^{-\frac{i q E e^{2 a x_L}}{4 a^2}}\frac{\Gamma(-\lambda_s^* + \nu^*)}{\Gamma\left(-\lambda_s^*+1\right) \Gamma (\nu^*)} H_s(y)\epsilon_s
\end{equation}
\begin{equation}
\label{fullmode11ILlim}
    V_{s,n}(x)_{\mathit{I^-_L}}=\frac{1}{R_s\;e^{\frac{a x_L}{2}}}\bigg(ie^\mu_a \gamma^{(a)}\partial_\mu-q A_\mu e^\mu_a \gamma^{(a)}+m\bigg)e^{i \omega (t_L+x_L)} e^{i k_z z} e^{-a x_L} e^{-\frac{i q E e^{2 a x_L}}{4 a^2}}\frac{\Gamma(-\lambda_s^* + 1)}{\Gamma(-\lambda_s^*+\nu^*) \Gamma (2-\nu^*)} H_s(y)\epsilon_s
\end{equation}
\bibliographystyle{cas-model2-names}

\bibliography{cas-refs}


\end{document}